\documentclass[fleqn,twoside,twocolumn,nofootinbib,showkeys,showpacs]{revtex4} % Specifies the document class %,unsortedaddress
\usepackage{amsmath,amssymb,amsthm}
\newtheorem{theor}{Theorem}%[section]
\theoremstyle{definition}
\newtheorem{cor}[theor]{Corollary}%[section]
\newtheorem{conjecture}[theor]{Conjecture}%[section]
\newtheorem{define}{Definition}%[section]
\newtheorem{example}{Example}%[section]
\theoremstyle{remark}
\newtheorem{rem}{Remark}%[section]
\newcommand{\BBS}{\mathbb{S}}%\newcommand{\BBT}{\mathbb{T}}
\newcommand{\BBE}{\mathbb{E}}
\newcommand{\cH}{\mathcal{H}}

\newcommand{\bx}{{\boldsymbol{x}}}
\newcommand{\Id}{{\mathrm d}}
\newcommand{\vx}{{\vec{\mathrm{x}}}}
\newcommand{\nC}{{\text{\textup{nC}}}}
\begin{document}
\title[Towards the noncommutative geometry of fundamental interactions]{TOWARDS AN AXIOMATIC MODEL OF FUNDAMENTAL INTERACTIONS\\ AT PLANCK SCALE}
\author{Arthemy V.~Kiselev}
\affiliation{Johann Ber\-nou\-lli Institute for Mathematics and Computer Science, University of Groningen,\\ %}
%\address{
P.O.~Box 407, 9700~AK Groningen, The Netherlands}
\email{A.V.Kiselev@rug.nl}

 %\razd{\seci} %1. Fields and elementary particles

\setcounter{page}{1}%
\setcounter{footnote}{1}%

\begin{abstract}\noindent%
By exploring %a 
possible physical sense %realisation 
of notions, structures, and logic in a class of noncommutative geometries, we try to unify the four fundamental interactions within an axiomatic quantum picture. We identify the objects and algebraic operations which could properly encode the formation and structure of sub\/-\/atomic particles, antimatter, annihilation, CP\/-\/symmetry violation, %mass and 
mass endowment mechanism, three lepton\/-\/neutrino matchings, spin, helicity and chirality, electric charge and electromagnetism, as well as the %short\/-\/range
weak and strong interaction between particles, admissible transition mechanisms \textup{(}e.g., %muon to muon neutrino, electron, and electron antineutrino
$\mu^{-}\mapsto\nu_\mu+\mathsf{e}^{-}+\overline{\nu}_{\mathsf{e}}$\textup{),} and decays \textup{(}e.g.,
%neutron to proton, electron, and electron antineutrino
$\mathsf{n}^0\mapsto\mathsf{p}^{+}+\mathsf{e}^{-}+\overline{\nu}_{\mathsf{e}}$\textup{).}
\end{abstract}


\maketitle

%\section{Introduction}\noindent%
\noindent\textbf{Introduction}\\[4pt]
In this paper we further our %earlier 
approach~\cite{SQS11,Protaras12} to an axiomatic description of the four fundamental interactions; for this we explore a natural class of noncommutative geometries~\cite{KontsevichCyclic,OlverSokolovCMP1998,Lorentz12}. Our approach is to some extent non\/-\/orthodox: we try to recognize physical phenomena beyond a given set of mathematical objects and structures; their algebraic simplicity is an evidence that such logic of information processing could be admissible or even dominant at Planck scale. Let us emphasize that we view quantum phenomena in the Universe as interaction of information codes so that the algebraic constructions at hand \emph{are} the objects of Nature.

This review is a sequel to~\cite{Protaras12}, where %in which 
we started to build the noncommutative model by considering a quasicrystal tiling of empty space. This tiling would yield an untwisted affine Lie algebra determined by the irreducible root system~$A_3$,\ $B_3$,\ or~$C_3$, see~\cite{KacInfDim}~--- if there were no defects in the crystal structure; note that the usefullness of the compactified fourth dimension which is associated by tadpoles~$\BBS^{\pm1}$ with the algebra's null vector will be revealed in section~\ref{SecCharge} of this paper.
We interpreted the time as a process of local topological reconfigurations of the lattice --\,whence the defects do emerge\,-- and we argued that contractions of edges in the $1$-\/skeleton of the cell complex at hand shows up as the mass.
The reverse process of spontaneous edge decontractions releases energy via $E=mc^2$. We thus noted that Hubble's law reflects a steady self\/-\/generation of space, whence we conjectured that the emission of cosmic microwave background radiation (CMBR) is an immanent property of space as topological manifold. This implies that the CMBR %radiation
cannot be altogether shielded, which suggests a possible (dis)\/verifying experiment checking our concept. By identifying the mass with reconfigurations in the topology of space, we revealed a candidate for the graviting but invisible `dark matter'~--- in this picture, it appears not as any form of matter but as a specific, meta\/-\/stable state of vacuum. 
(Another algebraic notion of (synonyms to) zero\/-\/length cyclic words shows up as a scalar field of `dark energy,' see section~\ref{SecPart} below.) In this paper we focus on the formation of matter and on the geometry of fundamental interactions in quantum space; we shall analyse the algebraic structures and logical operations which are immanent to that noncommutative world.

It must be noted that there co\/-\/exist many approaches to discretisation of the space\/-\/time and making it noncommutative; to the best of our knowledge, the paper~\cite{Wilson1974}    %lattice of discretely spaced points.
was seminal (see also~\cite{Creutz}). %for a pedagogical exposition.
  %after a long period of oblivion in the study of actual infinitesimally small numbers by Newton, Leibniz, and their successors.
A remarkable sample of axiomatisation trend is~\cite{Kapustin}.
Nowadays, various directions are represented, e.g., by~\cite{Connes,ConnesEtAl,DimitrijevicKulish,DouglasNekrasov,Ashtekar2003,EVerlinde2010}, etc.;
in contrast to \textit{loc.\ cit.}, the non\/-\/standard construction in~\cite{Atiyah2012} is built on an explicit assumption of sufficient smoothness for all structures at hand. However, it is readily seen that a benefit of space discretisation is gained from the use of finite, lattice\/-\/dependent adjacency tables for points which mark the quantum domains. This makes the loop quantum gravity paradigm~\cite{GambiniPullinQG} close to ours; yet we add the compactified, null dimension of the tadpoles~$\BBS^{\pm1}$, which relates the picture to Kaluza\/--\/Klein models. %[Rumer]
Continuing a comparison of our reasoning with existing schemes, let us recall that in~\cite{Protaras12} we analysed why a smooth, large\/-\/scale limit of the robust topological setup looks like the electroweak $U(1)\times SU(2)$-\/gauge theory. Furthermore, by introducing the noncommutative jet spaces %bundles 
and by re\/-\/establishing the entire calculus of variational multivectors in~\cite{SQS11,Lorentz12,GDE} we aim to properly encode the quarks, that is, confined building blocks of the strong interaction.
Finally, we recall that by a use of cyclic word approach~\cite{KontsevichCyclic} for coding sub\/-\/atomic particles one realizes them via closed strings of symbols written around the circles; the calculus of such necklaces then essentially amounts to %the 
familiar construction of topological pair of pants $S^1\times S^1\to S^1$, cf.~\cite{GreenSchwarzWitten}.\\%
%that transforms two circles into one (or vice versa).
\centerline{\rule{1in}{0.7pt}}

\noindent%
We attempt to understand the Universe as an information processor. While we only hypothesise about the algorithms which are actually employed to encode and process information, still we ought to define what information \emph{is} ---~at least in this context. Let us temporarily accept the following heuristic definition which appeals to a common\/-\/sense idea that information must be \emph{meaningful} and in principle \emph{verifiable}; hence our formula is non\/-\/rigorous, possibly incomplete or self\/-\/referring, and maybe contradictory; we also remark that the quantitative \emph{measurement} of information is a much more delicate issue (here we refer to the classical concepts of A.~N.~Kolmogorov, C.~E.~Shan\-non, etc.).

\begin{define}
Information is a rule %machine, automaton, black box
that inputs a message, that is, a sequence of $0$'s and $1$'s, and states a (non-)\/strict preference~$\preccurlyeq$ or~$\succcurlyeq$; by its output the rule attempts to predict the elementary event for the {next} digit of the message to be either~$0$ or~$1$. The non\/-\/strict preference $0\succcurlyeq1$ says that $0$~is not less likely than~$1$ whereas $0\preccurlyeq1$ tells us that $1$~is not less likely than~$0$. The act of comparison of the rule's prediction and the actually available next elementary event (possibly, itself being the first in a longer sequence) is a non\/-\/obligatory act of verification.

We accept that the processing of information is \emph{serious}: the rule states the {same} output on the same input data whenever an input is processed twice. If the rule exhibits its readiness to change opinion for repeated input, then the elementary events~$0$ and~$1$ are \emph{equally possible} (for example,
%\footnote{Likewise, an (in)\/finite sequence of recorded results for throwing a coin communicates the information that the coin was indeed thrown (in)\/finitely many times and, in the infinite case, it is likely that the outcome $0$ or $1$ of the next throwing are equally possible.
%  Thus, the only information which can be contained in the white noise (itself non\/-\/existent due to energy limitations) is that it indeed \textsl{is} white noise.}
an oracle is asked about the result of throwing a coin outside the light cone of its past). The non\/-\/strictness of preference builds the idea of uncertainty into information messages; however, if the rule states an absolute preference~$0\gg1$ or~$0\ll1$ for zero and unit, respectively, then the information is called \emph{precise} (yet it may be \emph{false}; uncertain information can also be true or false).
\end{define}

A sample application of such information messages is a check whether a given (i.e., contained in the input) object belongs to an encoded class of objects possessing a given property. For example, such is the coding of a triangle in a graph (see Fig.~\ref{FigLeptonNeutrino} on p.~\pageref{FigLeptonNeutrino}): whenever two sides are given, the rule discards all offered objects except for edges and all test edges except for the one which closes the contour.

On the other hand, a sample precise message ``The undetectable does exist''
affirmatively states --apart from the by\/-\/product information that there exist messages in general and there are means to encode and transmit them-- the existence of existence and perhaps the existence of the one who -- or something which created that message; other aspects of this message's meaning are non\/-\/verifiable hence non\/-\/informative.

Summarising, \emph{information} is (1.1) a formalised input and (1.2) a rule that establishes a preference for the next elementary event. In turn, \emph{processing} information is (2.1) the information itself and (2.2) a second\/-\/order rule that states a preference for the choice of new rules on the basis of input rules. For example, the decay of a free neutron is processing the rule that said `yes' whenever it was asked whether there was a neutron and which described that neutron; the processor's output states a preference that the new rules should ascertain the existence and describe the (motion of) proton, electron, and electron's antineutrino, see~\eqref{EqBetaDecay} on p.~\pageref{EqBetaDecay}.

The \emph{laws of Nature} are rules of third order, consisting of (3.1) second\/-\/order (reaction-) rules and (3.2) the rules to balance or modify the former (e.g., by prescribing the relative velocities). Such are the conservation laws for electric charge, energy, momentum, or angular momentum, etc. Notice that the laws of Nature do not refer to the formalised input~(1.1) by using which one encodes the actual presence and configuration of events or particles and their properties; these laws are universal.

It is perhaps appropriate to say that the set of \emph{fourth}\/-\/order rules for (non-)\/modification of the laws of Nature is a choice of the Universe itself. Indeed, should the fine\/-\/structure constant $\alpha=e^2/(\hbar c)\approx 1/137$ be varied or the dimension and topological properties of space (such as orientation or orientability) be changed, this would produce a different Universe; likewise, a slow modification of parameters in the laws of Nature (e.g., a drift of Hubble's constant in time or a change of half\/-\/life time for neutrons) would mean that the Universe itself is changing.

%\section*
\bigskip\noindent\textbf{Part~I. Noncommutative geometry of particles}

\smallskip\noindent%
For consistency, let us briefly recall the construction of quantum space; we refer to~\cite{Protaras12} for motivation and detail. To discretise space, we first consider its filling with a CW-\/complex; one may think that within sufficiently large domains --\,``large'' with respect to a count of edges, $\text{diam}\gg1$, along the $1$-\/skeleton\,-- the topology of such graph is that of crystal structure produced from the root systems~$A_3$,\ $B_3$,\ or~$C_3$;
their irreducibility prevents a slicing of space to lower\/-\/dimensional components. The choice of a root system from the set of these three is not definite so that superimposing local fragments of the respective CW-\/complexes  co\/-\/exist; a transition between the adjacency tables for different systems means a difference in organisation of information channels (see sec.~\ref{SecOscillations} below). At the walls of domains with regular crystal filling, the near\/-\/by vertices are interconnected by edges in a less regular way.

The cell complex which is \emph{dual} to the CW-\/complex at hand creates --\,by a Vo\-ro\-no\"\i\ tiling of space\,-- a quantum domain around each vertex in the old graph; the adjacency table for vertices becomes the table of neighbours for new domains. By construction, the old vertices act now as markers of quantum domains. In~\cite{Protaras12} we analysed the %several 
subtle aspects of taking continuous limits of such tilings by using an infinite bisection of edges.
Let us recall also that a macroscopic notion of length is not defined in such topological, \textsf{homeo}\/-\/class geometry; the distance between points and its measurement make sense only at large scale in the smooth, \textsf{diffeo}\/-\/class realisation of the Universe.

The reconfigurations of a given lattice consist of elementary events of edge (de)\/contractions, in the course of which the adjacency tables for a pair of neighbouring vertices are merged (so that the edge connecting them is destroyed) and the tadpoles, of which we speak in the next paragraph, are identified; the decontraction means: splitting of the table, insertion of the edge, and separation of the tadpoles. We noted that a count of faces for those cells in the dual complex which encapsulate the contraction defects could render the entropic origin of gravity (in conformity with~\cite{EVerlinde2010}). At the same time, we shall put the instability of particles in the context of spontaneous modifications in the topology of space and Hubble's~law.

\smallskip
The other step in the construction of quantum space is an attachment of tadpole to every vertex of the CW-\/complex' $1$-\/skeleton; in other words, the marker of each quantum domain is endowed with an extra edge which starts and ends there, going within the fourth, compactified dimension.

Let us suppose that the resulting CW-\/complex is oriented (cf.\ sec.~\ref{SecMatterVsAnti} below) so that $\BBS^{+1}$~is the path which runs along a tadpole in positive direction and $\BBS^{-1}$ is a walk against the wind. Denote by~$\vx_i$ the (vertex\/-\/dependent) collection of edges issued from a given vertex, and set~$\vx_i^{\,-1}$ for the same edges whenever they are went in the reverse direction whence %so that
$%\underrightarrow
{\vx_i\vx_i^{\,-1}}=1$ is the null path. We postulate that spatial edges are \emph{fermionic}: no path can run along an edge in the same direction twice. On the other hand, the tadpoles~$\BBS^{\pm1}$ are bosonic (why\,? see~\cite[Note~1]{Protaras12}). It is clear now that words written in the alphabet $\BBS^{\pm1}$,\ $\vx_i^{\,\pm1}$ encode paths, or walks along the lattice's edges.

\section{Formation and structure of sub\/-\/atomic particles}\label{SecPart}\noindent%
Suppose that the graph, i.e., $1$-\/skeleton of the CW\/-\/complex is given and $\BBS^{\pm1}$, $\vx_i^{\,\pm1}$ is the alphabet associated with its vertices;
we note that it can be vertex\/-\/dependent due to the on\/-\/going spontaneous splitting of edges, which creates the irregularity of the graph's structure and its possible local deviation from the regular tiling given by root systems.
   %\marginpar{Irreg $\neq$ contract}
Then every word in the alphabet(s) determines the path along the edges, starting at a given point; we shall consider the words of length zero and their synonyms
separately at the end of this section.

In what follows, we view paths as equivalence classes of walks because of a possibility to insert, wherever possible, trivial paths~$\langle a\cdot a^{-1}\rangle$ along the spatial component of the $1$-\/skeleton or a trivial word $\langle\BBS^1\cdot\BBS^{-1}\rangle=\langle1\rangle=\langle\BBS^{-1}\cdot\BBS^1\rangle$ by using the tadpole attached to every vertex. We %only 
recall that the spatial edges of the graph are fermionic and thus may be passed at most once in each of the two directions; this eliminates the risk of infinite loops.

We note that the paths (or walks) of positive proper length\footnote{The proper length of a path is the minimal number of edges in this path and in all its synonyms that may differ from it by synonyms of zero\/-\/length word inserted at any vertex along the way.} can happen to be closed, i.e., end at their starting point but not retract to it if one pulls by both ends of a thread that has been unrolled along the edges of the path. Notice further that such cycles can equivalently start and end at any other vertex along the contour; thus, the words encoding them are \emph{cyclic\/-\/invariant} (see~\cite{SQS11,Lorentz12} and~\cite{KontsevichCyclic}).

There are several mechanisms for a given path to be closed (apart from being a tadpole~$\BBS^{\pm1}$ hence closed by definition). 
First, there is a \emph{glossary}, i.e., a list of cyclic\/-\/invariant words (more precisely, a point\/-\/dependent \emph{gallery} of drawn contours); we shall quote from that source in sec.~\ref{SecPartExamples} and in Part~II from p.~\pageref{SecInter} onwards. 
Second, there could be an additional list of (formal sums of) paths which  themselves are not closed but which link to a contour whenever attached consecutively in suitable order, see sec.~\ref{SecInterStrong}. 
Thirdly, one can proclaim that a given path is closed by manually contracting a set of edges between its loose ends; this may require a considerable or infinite energy (recall that the topology of the CW\/-\/complex is trivial and one may not return to the starting point by walking in one direction and still coming back around the entire Universe). 
%Note that a single edge~$\vx_i$ in space can not be observed because the contraction of any path connecting its ends --and not being the reverse of~$\vx_i$, which would mean a null path,-- creates a tadpole, that is, just a point of space.

\begin{figure}[htb]
{\unitlength=1mm
\centerline{
\begin{picture}(75,45)%(0,-10)
%%%%%%%%%%%%%%%%%%%%%%%%%%%%%%%%%%%% OLD %%%%%%%%%%%%%%%%%%%%%
\put(0,37){
\begin{picture}(0,0)
\put(0,0){\circle*{2}}
\put(4,-1){$=$}
\put(10,0){\circle*{2}}
\put(10,0){\line(1,0){15}}
\put(20.5,2){\vector(-1,0){6}}
\put(14.5,-2){\vector(1,0){6}}
\put(25,0){\circle*{2}}
\put(29,-1){$=$}
\put(35,0){\circle*{2}}
\put(35,0){\line(1,0){10}}
\put(43,2){\vector(-1,0){6}}
\put(37,-2){\vector(1,0){6}}
\put(45,0){\line(1,1){5}}
\put(45,0){\line(1,-1){5}}
\put(50,5){\circle*{2}}
\put(50,-5){\circle*{2}}
\qbezier[12](50,5)(50,-5)(50,-5)
%%%%%%%%%%%%%%%%%%%%%%%%%%%%%%%%
\put(60,-1){$\Longrightarrow$}
%%%%%%%%%%%%%%%%%%%%%%%%%%%%%%%%
\end{picture}}
%%%%%%%%%%%%%%%%%%%%%%%%%%%%%%%%%%%%%%%%%%%
\put(20,12){\begin{picture}(0,0)(75,0)
\put(65,-1){\llap{$\Longrightarrow$}}
%%%%%%%%%%%%%%%%%%%%%%%%%
\put(75,0){\circle*{2}}
\put(75,0){\line(5,3){15}}
\put(75,0){\line(5,-3){15}}
\put(90,10){\circle*{2}}
\put(90,-10){\circle*{2}}
\put(90,10){\line(0,-1){20}}
\put(83,-1){$\circlearrowleft$}
\put(85,8){\vector(-4,-3){5}}
\put(80,-4.75){\vector(4,-3){5}}
%%%%%%%%%%%%%%%%%%%%%%%%%%%%%%%%
\put(100,-1){$\Longrightarrow$}
%%%%%%%%%%%%%%%%%%%%%%%%%%%%%%%%
\put(115,0){\circle*{2}}
\put(115,0){\line(5,3){15}}
\put(115,0){\line(5,-3){15}}
\put(130,-10){\vector(0,1){19}}
\put(116.5,1){\vector(-2,-1){0.5}}
\put(129,-8.5){\vector(2,-1){0.5}}
%\put(135,10){\vector(-2,-1){19.5}}
%\put(115,0){\vector(2,-1){19.5}}
%\put(135,-10){\vector(0,1){19.5}}
\put(130,-10){\circle*{2}}
\put(130,10){\circle*{2}}
\put(124,-1){$\circlearrowleft$}
\end{picture}}
%%%%%%%%%%%%%%%%%%%%%%%%%%%%%%%%%%%% OLD %%%%%%%%%%%%%%%%%%%%%
\end{picture}
}
}\caption{Creation of a contour.}\label{FigSorbonne}
\end{figure}
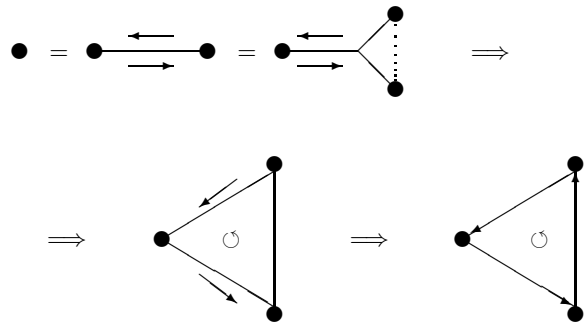

Next, there is a mechanism that generates cycles in the course of decontraction of edges.
  %so that a self\/-\/growing logos is immanent to phenomena of Nature.
Namely, consider the synonyms $\langle1\rangle=\langle\underrightarrow{\vx_{i\mathstrut}\vx_i^{\,-1}}\rangle$ and suppose that the vertex which the null path does not leave is a pair of vertices connected by a contracted edge, see Fig.~\ref{FigSorbonne}.
The formation of cycle is completed by matching the direction in which the new, decontracted edge is passed with the new face's orientation induced from the oriented CW\/-\/complex; note that energy is released in this process.

Finally, the energy\/-\/consuming scenario of cycle formation is an emission of \emph{two} closed contours which are walked in the opposite directions (see Fig.~\ref{FigNeutron}(b) on p.~\pageref{FigNeutron}): the there\/-\/and\/-\/back\/-\/again path visits its starting point thrice and is torn exactly in the middle, which creates two mutually inverse replicas; energy is spent on the disruption of the synonym of trivial word~$\langle1\rangle=\langle a\cdot a^{-1}\rangle$ in two nontrivial cyclic words~$\langle a\rangle$ and~$\langle a^{-1}\rangle$.

We postulate that \emph{particles} are the meaning of cyclic words that encode contours along the graph; such words contain the mark\/-\/up of contracted edges (this has nothing to do with the contractions con\-fi\-gu\-ra\-ti\-on for edges where the path does not run). 
We have noted in~\cite{Protaras12} %the previous section 
that the dynamics of contraction configurations determines the evolution of curvature and hence mass\/-\/energy.
We let %postulate that 
this be the mass endowment mechanism for %sub\/-\/atomic
particles: %, whose definition is introduced in the next section.
if there is at least one contracted edge along the contour, the particle is \emph{massive}; otherwise it is \emph{massless} (see Fig.~\ref{FigLeptonNeutrino} on p.~\pageref{FigLeptonNeutrino}).
Note that a given continuous contour with its contractions mark\/-\/up could have %in principle 
different masses with respect to the continuous limits of different tilings of space.

To formalise this approach in algebraic terms and make applicable the formalism of~\cite{Lorentz12}, we say that a {particle} is a functional
\[
\cH=\sum_{\langle{\text{words}}_{\bx}\rangle} \int %\veps(\bx,t)\cdot
\langle{\text{cycle}}_{\bx}\bigl(\BBS^{\pm1},\vx_i^{\,\pm}\bigr)\rangle
\,\Id\mu(\bx),
\]
where we use the following notation: The cycle is a closed contour that starts and ends at a point~$\bx$ of \emph{continuous} space (a reference of the particle to a point~$\bx$ is equivalent to referring it to any other point~$\boldsymbol{y}$ which lies on the contour passing through~$\bx$; still this amounts to a replacement of the contour's cyclic word by its equivalent,  starting the walk now at~$\boldsymbol{y}$, which leaves the contour intact); the {measure}, which refers to sets of points in space but does not exploit the notion of length, allows us to refer the particle to just one point -- or create a cloud of matter by spreading the contour in space over a given set;\footnote{A typical macroscopic diameter of such set of reference points would be the diameter of proton, which is $\approx 10^{-15}$\,\textsf{m}, making $\sim 10^{18}$ Planck units.} the units of measurement for the functional~$\cH$ are those of energy.

\begin{rem}
In terms of the \textsf{diffeo}\/-\/class geometry of~\cite{Lorentz12}, the functionals~$\cH=\sum_i\int h(\bx,t)$ $\langle{\text{word}}_i\rangle\,\Id\bx$ 
are {Hamiltonians} defined on the 
noncommutative space of maps from space to a free associative algebra's
quotient over relation of linear equivalence %relation
under cyclic permutations,
   %total space of the noncommutative tangent bundle over space, 
or on the total space of %the 
infinite jet bundle over such %bundle
space of maps. The {values} of such functionals are (formal sums of) possibly massive contours; particles interact by using algorithms and structures which we discuss in %sec.~\ref{SecInter}
Part~II (see also~\cite{Lorentz12}).
 %and~\cite{KontsevichFormality} for a \textsf{diffeo}\/-\/class, commutative version of the setup). 
Interactions between particles form a chain of events which contributes to a pace of time for a local observer.
\end{rem}

If a particle is referred to only one point of space, the discrete measure realises Dirac's $\delta$-\/distribution; note that it is the information about the contour and its properties which is ascribed to one point~--- still nothing is ``compressed.'' In particular, it is impossible to split an electron in  fragments and assemble it by trasporting such fragments from the spatial infinity to a given point, spending %an 
infinite energy to overcome the repulsion potential of the would\/-\/be fractions of the charge~$-e$; this approach also resolves the difficulty with an infinite density of electron's mass. Indeed, the particle is proclaimed existing at a certain point of space.

On the other hand, for dimensionful particles like proton or neutron the measure is concentrated on a larger set; typically, its support is a connected bounded set. We thus exclude from further consideration rapidly descreasing distributions with --juridically speaking, unbounded-- supports.

\begin{rem}
The interactions operate with the \emph{values} of the functionals that encode particles but not with the functionals themselves; in effect, particles interact as indivisible entities so that the processes refer to the particles' existence but not to their instant ``shapes,'' for those are undefined due to the Heisenberg uncertainty principle. For example, in the course of decay $\mathsf{n}^0\longmapsto\mathsf{p}^{+}+\mathsf{e}^{-}+\overline{\nu}_{\mathsf{e}}$ the free neutron stops existing in space at all its points when proton and the other two particles are formed or start to form.
\end{rem}
 %therefore, for the sake of brevity we shall 

\begin{rem}
The edges which are contracted break the symmetry of any contour; they also produce the irregularity of the graph's structure near the merging vertices, not necessarily at %the 
points of the contour. Such defects induce the particle 
to interact with other objects; conversely, massless point particles with simple contours without contracted edges or marked vertices (e.g., not carrying electric charge) demonstrate very low cross\/-\/sections for interaction with matter.
\end{rem}

%\begin{rem}Inertial mass grows with relative velocity.
%\end{rem}

The scalar field of {zero\/-\/length} words is the density of vacuum energy. It does not show up in the form of energy communicated to any particles chiefly because of absence of those particles; it just is. 
%The sum of the vacuum energy values assigned to the vertices in any finite domain is finite. %versus the self-generation of space
However, we already know that the trivial word~$\langle1\rangle$ is synonymic to paths~$\langle a\cdot a^{-1}\rangle$ walked twice, there and back again; in particular, it is synonymic to a closed cycle~$a$ and then~$a^{-1}$ walked in the opposite directions. By spending some extra energy on disrupting the contour~$a$ from the anticontour~$a^{-1}$, which creates two cyclic\/-\/invariant words of positive length, we convert a part of the spare vacuum energy into the \emph{matter\/-\/antimatter} %spontaneous
pair of particles.

\section{Matter versus antimatter}\label{SecMatterVsAnti}\noindent%
The orientation of the CW-\/complex distinguishes between two directions to walk around a given contour (e.g., a face of a cell).\footnote{We emphasize that the order in which one passes the edges when reading the contour's cyclic word is not the same as a choice of the contour's orientation (either matching or reverse with respect to the orientation induced from the CW-\/complex); we reserve that choice for the definition of spin.} To pass a contour backwards, the relay is this:
\begin{itemize}
\item replace each letter $\BBS^{\pm1}$ or $\vx_i^{\,\pm1}$ with its inverse, resp., $\BBS^{\mp1}$ and $\vx_i^{\,\mp1}$;
\item read the word backwards, i.e., in the right\/-\/to\/-\/left order.
\end{itemize}
This mechanism tells matter from antimatter; the same principle is applicable literally to formal sums of non\/-\/closed paths, see sec.~\ref{SecInterStrong}.
Thus, a distinction between matter and antimatter is conventional; yet the two anti\/-\/worlds can differ in their physical properties due to the CP\/-\/symmetry violation (see below).

%\begin{convention}
%A positive motion along the 
%\end{convention}
%\marginpar{Convent}
Let us notice further that by pasting a word that means matter at the beginning or at the end of the respective word for antimatter, or \textit{vice versa}, one obtains the trivial product~$\langle1\rangle$ of the two paths. This is why energy is released in the course of \emph{annihilation}; such energy impulse can take the shape of a photon\/-\/antiphoton pair, etc.\ (see sec.~\ref{SecCharge}); after its minor part is spent on the disruption of contours, the exact amount of released energy depends on the mass\/-\/energy of the two vanishing enantomorphs.

The fact that the CW-\/complex is oriented not only distinguishes between Left~(L) and Right~(R) but also motivates a possible violation of the CP-\/symmetry, which itself is the mirror\/-\/reflection Left~$\rightleftarrows$ Right in the orientation of space \emph{and} a substitution matter~$\rightleftarrows$ antimatter (i.e., reading backwards all the words from the glossary).

\begin{proof}[Explanation]
Let us take a --now\/-\/existing-- tetrahedron in the spatial part of the graph and contract it along three bold edges as in Fig.~\ref{FigTetra}, doing this in two mirror\/-\/reflected ways (note that 
the orientation swap Left~$\rightleftarrows$ Right is \emph{local} so that  space is not turned inside\/-\/out; the labels of vertices remain what they are because it is only the edge contraction scenario which is reflected in the mirror).
  %marking the vertices by digits is an act of will yet the two scenarios are true mirror copies of each other).
%%%%%%%%%%%%%%%%%%%%%
%%% Figure 2.  
%%%%%%%%%%%%%%%%%%%%%
Let us recall now that a catalogued particle itself and the processor which handles particles --e.g., by disrupting contours and reconfiguring the available edges-- is an automaton: it reads the (cyclic) words from the glossary\footnote{The catalogue of matter and antimatter is the glossary of equivalence classes of contour\/-\/determining cyclic words with a mark\/-\/up of the edges to\/-\/contract; the glossary is independent from the actual configuration of contractions in the graph~--- it is indeed a list of words.}
and crawls along the contracted graph by running a program like this:
\begin{description}
\item[\textmd{R: }] \texttt{move right, then right, then right again} 
\end{description}
on the plane containing Fig.~\ref{FigTetra}(d). Under the instant CP-\/transformation, the automaton is interrupted and starts the relay program\footnote{Notice that the order of reading letters in catalogued words and the arising precedence ``before'' and antecedence ``after'' have nothing to do with the time as physical process; it is not appropriate to postulate that antimatter flies backwards in time.}
\begin{description}
\item[\textmd{L: }] \texttt{move left, but before that move left, still before which move left}
\end{description}
on the plane in Fig.~\ref{FigTetra}(a).
  
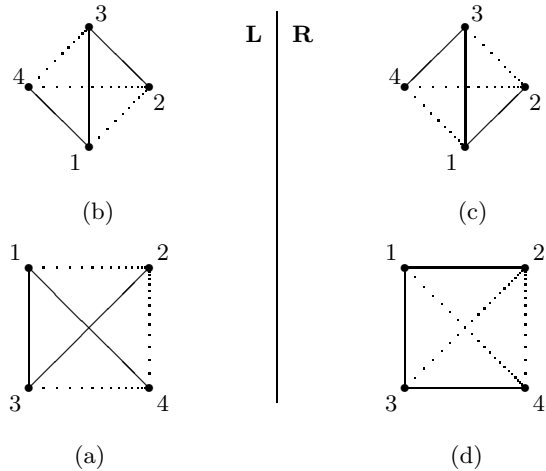
\begin{figure}[htb]
{\unitlength=1mm
\centerline{\begin{picture}(70,60)
%%%%%%%%%%%%%%%%%%%%%%%%%%%%%%%%%%%%%  OLD  %%%%%%%%%%%%%%%%%%%%%%%
\put(0,9){\begin{picture}(16,16)
\put(0,0){\circle*{1}}
\put(16,0){\circle*{1}}
\put(0,16){\circle*{1}}
\put(16,16){\circle*{1}}
\put(0,0){\line(0,1){16}}
\put(0,0){\line(1,1){16}}
\put(0,16){\line(1,-1){16}}
\qbezier[16](0,0)(16,0)(16,0)
\qbezier[16](0,16)(16,16)(16,16)
\qbezier[16](16,0)(16,16)(16,16)
\put(-1,-3){\llap{3}}
\put(17,-3){4}
\put(-1,17){\llap{1}}
\put(17,17){2}
\put(6,-10){(a)}
\end{picture}
}
%%%%%%%%%%%%%%%%%%%%%%%%%%%%%%%%%%%%%%%
\put(8,41){\begin{picture}(16,16)
\put(0,0){\circle*{1}}
\put(0,0){\line(0,1){16}}
\put(0,16){\circle*{1}}
\put(0,0){\line(-1,1){8}}
\put(-8,8){\circle*{1}}
\put(0,16){\line(1,-1){8}}
\put(8,8){\circle*{1}}
\qbezier[16](-8,8)(8,8)(8,8)
\qbezier[12](-8,8)(0,16)(0,16)
\qbezier[12](0,0)(8,8)(8,8)
\put(-1,-3){\llap{1}}
\put(8.5,5){2}
\put(0.8,16.8){3}
\put(-8.5,8.5){\llap{4}}
\put(-1,-9.5){(b)}
\end{picture}
}
%%%%%%%%%%%%%%%%%%%%%%%%%%%%%%%%%%%%%%%
\put(33,31){\begin{picture}(0,0)(75,0)
\put(73,24){\llap{\textbf{L}}}
%\put(75,6){\line(0,1){22}}
\put(75,-24){\line(0,1){52}}
\put(77,24){\textbf{R}}
\end{picture}}
%%%%%%%%%%%%%%%%%%%%%%%%%%%%%%%%%%%%%%%
%\put(33,-2){\begin{picture}(0,0)(75,0)
%\put(73,24){\llap{\textbf{L}}}
%\put(75,6){\line(0,1){22}}
%\put(77,24){\textbf{R}}
%\end{picture}}
%%%%%%%%%%%%%%%%%%%%%%%%%%%%%%%%%%%%%%%
\put(58,41){\begin{picture}(16,16)
\put(0,0){\circle*{1}}
\put(0,0){\line(0,1){16}}
\put(0,16){\circle*{1}}
\put(0,16){\line(-1,-1){8}}
\put(-8,8){\circle*{1}}
\put(0,0){\line(1,1){8}}
\put(8,8){\circle*{1}}
\qbezier[16](-8,8)(8,8)(8,8)
\qbezier[12](-8,8)(0,0)(0,0)
\qbezier[12](0,16)(8,8)(8,8)
\put(-1,-3){\llap{1}}
\put(8.5,5){2}
\put(0.8,16.8){3}
\put(-8.5,8.5){\llap{4}}
\put(-1,-9.5){(c)}
\end{picture}
}
%%%%%%%%%%%%%%%%%%%%%%%%%%%%%%%%%%%%%%%
\put(50,9){\begin{picture}(16,16)
\put(0,0){\circle*{1}}
\put(16,0){\circle*{1}}
\put(0,16){\circle*{1}}
\put(16,16){\circle*{1}}
\put(0,0){\line(0,1){16}}
\put(0,0){\line(1,0){16}}
\put(0,16){\line(1,0){16}}
\qbezier[24](0,0)(16,16)(16,16)
\qbezier[24](0,16)(16,0)(16,0)
\qbezier[16](16,0)(16,16)(16,16)
\put(-1,-3){\llap{3}}
\put(17,-3){4}
\put(-1,17){\llap{1}}
\put(17,17){2}
\put(6,-10){(d)}
\end{picture}
}
%%%%%%%%%%%%%%%%%%%%%%%%%%%%%%%%%%%%%  OLD  %%%%%%%%%%%%%%%%%%%%%%%
\end{picture}
}
}\caption{Mirror\/-\/reflected %symmetric 
contractions.}\label{FigTetra}
\end{figure}

Suppose for definition that the tetrahedra in Fig.~\ref{FigTetra}(b-c) are \emph{the only} contracted edges in the Universe and it is these two objects which encode the choice of its orientation. The mechanism of CP-\/symmetry violation is that the mirror\/-\/reflected %symmetric 
contractions of the tetrahedron produce unequal configurations of the tadpoles, see Fig.~\ref{FigTadpoles}.
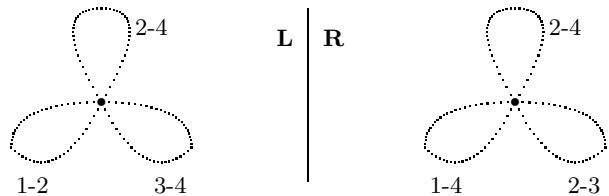
\begin{figure}[htb]
{\unitlength=1mm
\centerline{\begin{picture}(80,25)
\put(12.5,12.5){\begin{picture}(12.5,12.5)
\put(0,0){\circle*{1}}
\qbezier[24](0,0)(-7.5,12.5)(0,12.5)
\qbezier[24](0,0)(7.5,12.5)(0,12.5)
\qbezier[24](0,0)(-12,0)(-12,-6)
\qbezier[24](0,0)(-6,-12)(-12,-6)
\qbezier[24](0,0)(12,0)(12,-6)
\qbezier[24](0,0)(6,-12)(12,-6)
\put(4.5,9){2-4}
\put(-7,-12){\llap{1-2}}
\put(7,-12){3-4}
\end{picture}
}
%%%%%%%%%%%%%%%%%
\put(38,20){\llap{\textbf{L}}}
\put(40,2){\line(0,1){23}}
\put(42,20){\textbf{R}}
%%%%%%%%%%%%%%%%%
\put(67.5,12.5){\begin{picture}(12.5,12.5)
\put(0,0){\circle*{1}}
\qbezier[24](0,0)(-7.5,12.5)(0,12.5)
\qbezier[24](0,0)(7.5,12.5)(0,12.5)
\qbezier[24](0,0)(-12,0)(-12,-6)
\qbezier[24](0,0)(-6,-12)(-12,-6)
\qbezier[24](0,0)(12,0)(12,-6)
\qbezier[24](0,0)(6,-12)(12,-6)
\put(4.5,9){2-4}
\put(-7,-12){\llap{1-4}}
\put(7,-12){2-3}
\end{picture}
}
\end{picture}
}
}\caption{Formation of tadpoles in Fig.~\ref{FigTetra}.}\label{FigTadpoles}
\end{figure}
Thus, the R-\/automaton that runs the R-\/program and, for definition, reaches the vertex~$\underline{1}$ after its first step terminates the program at the point two steps to the right from the vertex~$\underline{2}$. Reshape now the tetrahedron's contraction and switch to the antiparticles and L-\/program; the L-\/automaton starts at the R-automaton's endpoint, runs the L-\/program, and terminates at the point~$\underline{1}$ (but \emph{not} at the starting point of the R-\/automaton) because the edge $\underline{2}\to\underline{1}$ becomes a tadpole in the antiworld instead of being contracted in the right\/-\/oriented world (the orientation is determined by the order $1\prec2\prec3\prec4$ in Fig.~\ref{FigTetra}(b)).

In particular, it may happen that by running its R-\/program the R-\/automaton crawled along the \emph{closed} contour while reading the right words from the glossary but the path of the L-\/automaton appears to be \emph{not closed}, after the orientation was reversed by switching from Fig.~\ref{FigTetra}(c) to Fig.~\ref{FigTetra}(b). Recall that a reaction is information\/-\/processing the output of which is a word, and the Left and Right automata attempt to create the particles' contours by reading this word in one of the two possible directions. We conclude that, specifically to the scenario which we have discussed, the L-\/channel either is suppressed altogether (by an earlier convention that spatial edges are fermionic) or is possible at an expense of energy needed to contract extra edge(s) in order to close the L-\/path and thus bring the L-\/automaton to its starting point.

This produces the CP-\/asymmetry if the preference of Right over Left is implanted in the Universe since the moment of decontraction of the first tetrahedron in its history.
\end{proof}

\begin{cor}
The {masses} of particles and respective antiparticles can be (slightly) unequal.
\end{cor}

\begin{cor}
If Right has been prevailing over Left since the moment when creation of particles became possible, there is an imbalance between matter and antimatter nowadays.
\end{cor}

\begin{rem}
The risk of CP-\/symmetry violation is built into the automaton which (1) reads from the glossary dogmatically and (2) attempts, by disrupting a word in between two letters and by having disrupted another given cyclic word, to paste that \emph{open} word from the glossary in between, or immediately before or after the letters of the given word, and then (3) attempts to realise the output as a route along the graph~--- not taking into account the actual configuration of contractions.

On the contrary, the functionality of a different\/-\/type automaton that handles already\/-\/existing closed contours in space is stable. Indeed, the already paved routes remain closed if they had this property before the CP-\/transformation %and were not disrupted at any vertex 
(note that the decontracting edges are pasted into the contours), see sec.~\ref{SecInterStrong} for details.
\end{rem}

\section{Properties and examples of elementary particles}\label{SecPartExamples}\noindent%
The definition of quantum numbers and each act of their measurement channels information between tiling(s) of the \textsf{homeo}\/-\/class topological manifold by a CW-\/complex and the \textsf{diffeo}\/-\/class, smooth and commutative visible space. There is no surprise in that the available data which can be transmitted in this way must be very rough and appeal to topological invariants only.

\subsection{The spin}\label{SecSpin}\noindent%
From now on let us suppose that the contours which determine particles are \emph{oriented}; note that a choice of orientation for the contour's edges within each of its loops (e.g., consider a bouquet of circles) is in general not correlated with the order in which these edges are passed when one reads the particle's cyclic word.\footnote{Experiments report that the electron neutrinos~$\nu_{\mathsf{e}}$ usually have \emph{left helicity}, that is, the orientation of their contours makes left\/-\/handed helix propagating in the direction of their macroscopic instant velocity; likewise, the electron \emph{anti}neutrinos~$\overline{\nu}_{\mathsf{e}}$ tend to have right helicity, see Fig.~\ref{FigLeptonNeutrino} on p.~\pageref{FigLeptonNeutrino}.}
We notice also that a contour's orientation is thus not necessarily inherited from the orientation of the CW\/-\/complex.

\begin{define}
The \emph{spin} of a given particle is a quantum number which is equal to the sum
\[
s={\sum_{\text{loops}}}\pm\tfrac{1}{2}\hbar
\]
over all the loops in the spatial component of the closed contour. (The tadpoles $\BBS^{\pm1}$ at each vertex are dealt with separately in the next section). The contribution of a loop is~$+\tfrac{1}{2}\hbar$ if the choice of its orientation coincides with the order in which these edges are written in the cyclic word, and equals~$-\tfrac{1}{2}\hbar$ otherwise.
\end{define}

Almost all known particles do have spin and most of the stable particles (either massive or without mass) are coded by simple contours --the unknots-- so that their spins are~$\pm\tfrac{1}{2}\hbar$.

\begin{rem}
The orientation of a loop in a contour can instantly switch from~$+\tfrac{1}{2}\hbar$ to~$-\tfrac{1}{2}\hbar$, making no effect on the orientation of other loops in the contour, so that the overall value of the spin changes by unit steps~$\pm\hbar$ between its minimal nonpositive and maximal nonnegative values.
\end{rem}

\begin{rem}
In view of a possibility for an sudden swap of orientation of a given loop, it is still meaningful to say that products of a weak reaction --in the course of which the contours are instantly disrupted and recombined-- can at that moment retain the orientation of edges from loops in the input particles. We therefore expect that the spins of free neutrons and protons which are emitted in the $\beta$-\/decay $\mathsf{n}^0\longmapsto\mathsf{p}^{+}+\mathsf{e}^{-}+\overline{\nu}_{\mathsf{e}}$ {are} correlated (though for none of them we can measure for certain the projection of spin to a given direction in the macroscopic space at the moment of decay, and though neither before nor after the reaction two measurements of the same particle's spin would necessarily coincide).
\end{rem}

\subsection{The electric charge}\label{SecCharge}\noindent%
Let us recall from~\cite{Protaras12} %sec.~\ref{SecSpace} 
that each vertex of the CW-\/complex (and every point in the continuous limit of a tiling) carries the %bosonic 
tadpole(s)~$\BBS^1$ %=\overrightarrow{\mathsf{tt}}
starting and ending at that point. 
%This makes the electromagnetism a very famous example of a cyclic\/-\/invariant theory.

\begin{define}
The \emph{electric charge} of a given particle is a quantum number which is equal to the difference
\[
c=\sum_{\langle\text{words}\rangle} \bigl(\sharp\,\BBS^1 - \sharp\,\BBS^{-1}\bigr)\cdot e
\]
of the numbers of time that particle's contour passes the tadpoles --as they are written in the particle's word(s)-- in positive and negative directions, each loop in the compactified dimension thus contributing with the respective elementary charge~$\pm e$.
\end{define}

\begin{cor}
The electric charges of particles and their antiparticles are equal by absolute value and have opposite signs.
\end{cor}

\begin{rem}
In the beginning, the initial electric charge of the Universe was equal to zero because the contours did not exist yet (for all the tadpoles were contracted).
\end{rem}

\begin{cor}
Nowadays, the Universe is overall electrically neutral because all %the
separately existing electric charges in it were obtained by disruption of neutral contours (indeed, all the contours were obtained from copies of the trivial word~$\langle1\rangle$ by decontraction of edges, see Fig.~\ref{FigSorbonne} on p.~\pageref{FigSorbonne}, or by disruptions, see Fig.~\ref{FigNeutron} on p.~\pageref{FigNeutron}).
\end{cor}

\begin{rem}
There is a temporary shortage of %the 
{magnetic} charges in this Universe.
\end{rem}

\subsection{The photon}\label{SecPhoton}\noindent%
By introducing the following definition we resolve a delicate issue that the photon is a bosonic particle that can exist in two possible states (or \emph{polarisations}), which usually requires that one manually removes the spin\/-\/zero state from the %spin 
triplet~$s\in\{-\hbar,0,\hbar\}$.

\begin{define}
The \emph{polarised photons}~$\gamma$ are massless point particles whose cyclic words are
\begin{align*}
\gamma_{\circlearrowleft}&=\langle\BBS^1_m\BBS^{-1}_n\rangle,\\
\gamma_{\circlearrowright}&=\langle\BBS^{-1}_m\BBS^1_n\rangle,\qquad m<n,
\end{align*}
where $m$,\ $n\in%\BBZ\subseteq
\mathcal{I}$ belong to the ordered indexing set at a point of quantum space
(see \cite[Note~1]{Protaras12});
photons carry no electric charge ($\pm e\mp e\equiv0$).
\end{define}

\begin{cor}
The polarised \emph{antiphotons}~$\overline{\gamma}$,
\begin{align*}
\overline{\gamma}_{\circlearrowleft}&=\underrightarrow{\langle\BBS^1_m\BBS^{-1}_n\rangle}^{-1}=\underrightarrow{\langle\BBS^1_n\BBS^{-1}_m\rangle}=\langle\BBS_m^{-1}\BBS_n^1\rangle=\gamma_{\circlearrowright},\\
\overline{\gamma}_{\circlearrowright}&=\underrightarrow{\langle\BBS_m^{-1}\BBS_n^1\rangle}^{-1}=\underrightarrow{\langle\BBS_n^{-1}\BBS_m^1\rangle}=
\langle\BBS_m^1\BBS_n^{-1}\rangle=\gamma_{\circlearrowleft},
\end{align*}
(here $m<n$) are identical to the photons with opposite polarisations.
\end{cor}

\begin{rem}
Photons travel\footnote{In~\cite{Protaras12} %sec.~\ref{SecTime} 
we understood the propagation of a photon as the automaton that creates unit events by destroying the photon at one end of an edge in the graph and creating the same photon at the other end. The count of such events is the pace of time.} 
in space with invariant light speed~$c$. However, when a photon is stopped by a material object in {continuous} space at a given point, the indexed set of fermionic tadpoles at that point reduces to a unique circle~$\BBS^{\pm1}$ so that the photon~$\langle\BBS^{\pm1}\BBS^{\mp1}\rangle$ immediately becomes synonymic to the zero\/-\/length word $\langle1\rangle$.
Thus, photons~$\gamma$ play the r\^o\-le of energy carriers.
\end{rem}

\subsection{Lepton\/--\/neutrino matchings}\label{SecOscillations}\noindent%
Let us use the tiling associated with the root system~$A_3$; this 
defines %prescribes 
the configuration of local information channels for the rules which process information about particles in the course of reactions.

\begin{define}
The \emph{electron}~$\mathsf{e}^{-}$, antielectron (or \emph{positron})~$\mathsf{e}^{+}$, \emph{electron neutrino}~$\nu_{\mathsf{e}}$, and \emph{electron antineutrino}~$\overline{\nu}_{\mathsf{e}}$ are point particles whose contours are drawn --schematically and up to homeomorphisms-- in Fig.~\ref{FigLeptonNeutrino}; the contracted edges are marked by dotted lines. %taking the respective letters in parentheses.
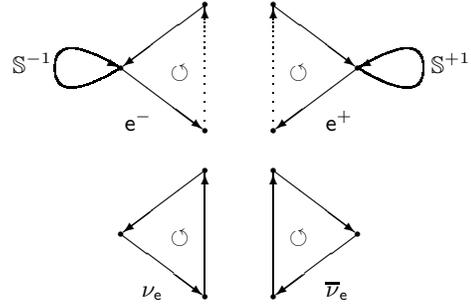
\begin{figure}[htb]
{\unitlength=0.7mm
\centerline{
\begin{picture}(70,53)
%%%%%%%%%%%%%%%%%%%%%%%%%%%%%%%%%% OLD %%%%%%%%%%%%%%%%%%%%%%%%%
\put(10,3){\begin{picture}(16,24)(0,-8)
\put(0,0){\circle*{1}}
\put(0,0){\vector(4,-3){15.5}}
\put(16,-12){\circle*{1}}
\put(16,-12){\vector(0,1){23.5}}
\put(16,12){\circle*{1}}
\put(16,12){\vector(-4,-3){15.5}}
\put(13,-2){\llap{$\circlearrowleft$}}
\put(8,-12){\llap{$\nu_{\mathsf{e}}$}}
\end{picture}
}
%%%%%%%%%%%%%%%%%%%%%%%%%
\put(10,34.5){\begin{picture}(16,24)(0,-8)
\put(0,0){\circle*{1}}
\put(0,0){\vector(4,-3){15.5}}
\put(16,-12){\circle*{1}}
\qbezier[24](16,-12)(16,12)(16,12)
\put(16,10.5){\vector(0,1){1}}
\put(16,12){\circle*{1}}
\put(16,12){\vector(-4,-3){15.5}}
\qbezier(0,0)(-12.5,7.5)(-12.5,0)
\qbezier(0,0)(-12.5,-7.5)(-12.5,0)
\put(-2,1){\vector(2,-1){1.5}}
\put(-13,-1){\llap{$\BBS^{-1}$}}
\put(13,-2){\llap{$\circlearrowleft$}}
\put(6,-12){\llap{$\mathsf{e}^{-}$}}
\end{picture}
}
%%%%%%%%%%%%%%%%%%%%%%%%%
\put(55,34.5){\begin{picture}(16,24)(0,-8)
\put(0,0){\circle*{1}}
\put(0,0){\vector(-4,-3){15.5}}
\put(-16,-12){\circle*{1}}
\qbezier[24](-16,-12)(-16,12)(-16,12)
\put(-16,10.5){\vector(0,1){1}}
\put(-16,12){\circle*{1}}
\put(-16,12){\vector(4,-3){15.5}}
\qbezier(0,0)(12.5,7.5)(12.5,0)
\qbezier(0,0)(12.5,-7.5)(12.5,0)
\put(2,1){\vector(-2,-1){1.5}}
\put(14,-1){$\BBS^{+1}$}
\put(-13,-2){$\circlearrowleft$}
\put(-6,-12){$\mathsf{e}^{+}$}
\end{picture}
}
%%%%%%%%%%%%%%%%%%%%%%%%%
\put(55,3){\begin{picture}(16,24)(0,-8)
\put(0,0){\circle*{1}}
\put(0,0){\vector(-4,-3){15.5}}
\put(-16,-12){\circle*{1}}
\put(-16,-12){\vector(0,1){23.5}}
\put(-16,12){\circle*{1}}
\put(-16,12){\vector(4,-3){15.5}}
\put(-13,-2){$\circlearrowleft$}
\put(-6,-12){$\overline{\nu}_{\mathsf{e}}$}
\end{picture}
}
%%%%%%%%%%%%%%%%%%%%%%%%%%%%%%%%  OLD  %%%%%%%%%%%%%%%%%%%%%%%%%
\end{picture}
}
}\caption{(Anti)electron and electron (anti)neutrino.}\label{FigLeptonNeutrino}
\end{figure}
Each of the four particles has spin $\pm\tfrac{1}{2}\hbar$ due to the two possibilities to choose an orientation of these unknots in~$\BBE^3$.
The electron~$\mathsf{e}^{-}$ has negative electric charge~${-}e$ and that of positron~$\mathsf{e}^{+}$ equals~${+}e$.
\end{define}

\begin{cor}
Apart from carrying no electric charge and probably having no mass, the electron (anti)\/neutrino is in all other respects ``indistinguishable'' from the (anti)\/electron.
\end{cor}

The mass of (anti)electron is app\-ro\-xi\-ma\-te\-ly $0.511$~\textsf{MeV}$/c^2$; by using Fig.~\ref{FigSorbonne} on p.~\pageref{FigSorbonne} we argued that there \emph{are} chargeless massless spin\/-\/$\tfrac{1}{2}\hbar$ particles and now we identify them with electron (anti)\/neutrinos; however, our reasoning does not forbid the existence of particles with very similar properties yet with a tiny mass (hence travelling slower than light). If being massless and thus having no naturally marked vertex or edge --unlike the (anti)\/electron-- neutrinos symptomise a general disinclination to interaction of any kind with any type of matter.

However, we owe almost everything to these simplest neutrinos because it was the contour~$\nu_e$ which was first created in the spatial component of the CW-\/complex in the course of decontraction of the first oriented face, see Fig.~\ref{FigSorbonne}; the first photon~$\gamma$ appeared at about the same time, after the decontraction of the first tadpole; we see no reason to debate which 
%event happened earlier.
word was first.\\
\centerline{\rule{1in}{0.7pt}}

\noindent%
The two heavier point leptons are the (anti)\emph{muon}~$\mu^{\pm}$ ($\mathsf{m}_\mu\approx 105.7$%\text
{ \textsf{MeV}}$/c^2$) and (anti)\emph{tau} $\tau^{\pm}$ ($\mathsf{m}_\tau\approx 1776.8$%\text
{ \textsf{MeV}}$/c^2$); these four particles are unstable (their proper %half\/-\/
life\/-\/times are approximately $2\,\mu\mathsf{s}$ and $3\cdot 10^{-13}$\,\textsf{s}, respectively); in the course of decay they are reported~\cite{Okun} to produce the respective (anti)neutrinos $\nu_\mu$,\ $\overline{\nu}_\mu$,\ $\nu_\tau$, and~$\overline{\nu}_\tau$, which is quite logical: after %a %the
decontraction of heavy leptons' contours, their electric charges are decoupled --by spending %a 
minor part of the released energy on  detachment of the charge tadpoles-- and re\/-\/attached to the newly\/-\/generated, then separated mutually inverse pairs of contours for lighter leptons and their antineutrinos.

\begin{example}
Consider the Michel decay $\mu^{-}\longmapsto \nu_\mu+\mathsf{e}^{-}+\overline{\nu}_{\mathsf{e}}$. The decontraction of edges which endow the muon~$\mu^{-}$ with mass and deprivation of its contour from its negative electric charge, which is encoded by the tadpole~$\BBS^{-1}$, produces the neutrino~$\nu_\mu$ and a store of dispensable energy, which has been partly used to detach the charge and which is used then to disrupt the trivial word issued from the point where the charge is located,
\[
\langle1\rangle=\langle\nu_{\mathsf{e}\mathstrut}\cdot{\nu}_{\mathsf{e}}^{\,-1}\rangle\longmapsto\langle\nu_{\mathsf{e}}\cdot\BBS^{-1}\rangle+\langle\overline{\nu}_{\mathsf{e}}\rangle.
\]
A contraction of edge(s) in the charged first term of the right\/-\/hand side yields the second and third particles in the reaction's output. The remainder of energy (left from the excess of muon's mass) is communicated to the three particles~$\nu_\mu$,\ $\mathsf{e}^{-}$, and~$\overline{\nu}_{\mathsf{e}}$ as their kinetic energy. (Note the overall conservation of %the 
electric charge and a likely preservation of %the 
muon's spin by the muon neutrino~$\nu_\mu$.)
\end{example}

However, we see that the discarded contour for~$\nu_\mu$ in Michel's reaction and the mutually reverse prototypes for~$\mathsf{e}^{-}$ and $\overline{\nu}_{\mathsf{e}}$ could co\/-\/exist on \emph{different} CW\/-\/complexes which fill in a unique continuous \textsf{homeo}\/-\/class Universe; we notice that the trivial word~$\langle1\rangle$ to\/-\/expand at the location of the charge does not refer to a choice of the alphabet, whereas the store of energy and the loop along~$\BBS^{-1}$, not leaving the point of physical space, are logically transportable between the schemes of information processing.

If so, the decays of the heaviest~$\tau^{\pm}$ and less heavy~$\mu^{\pm}$ could be second\/-\/order phase transitions, in the course of which the information channels are locally reconfigured (i.e., one crystal structure transforms into another --no less well ordered-- so that the order parameter is constant) and the energy and electric charge are pumped into the new logical processor of information; the old processor calculates the receding of muon's neutrino.

\begin{conjecture}
There \emph{are}, and there are exactly \emph{three} types of leptons --$\mathsf{e}$,\ $\mu$, and~$\tau$-- because there are exactly three types of irreducible lattices --$A_3$,\ $B_3$, and~$C_3$-- in Euclidean space~$\BBE^3$, open domains in which are homeomorphic to the spatial components of domains in \textsf{homeo}\/-\/class realisation of the Universe.
\end{conjecture}

%%%%%%%%%%%%%%%%%%%%%%%%%%%%%%%%%%%%%%%%%%%%%%%%%%%%%%
\bigskip\noindent\textbf{Part~II. Geometry of fundamental interactions}\label{SecInter}

\medskip\noindent%
Let us focus now on the algebraic structures of the four fundamental interactions; we assume for definition that the processes of reconfiguration and interaction occur on the same lattice so that the alphabet $\vx_i^{\,\pm1}$, $\mathbb{S}^{\pm1}$ %=\overrightarrow{\mathrm{tt}}$ 
is common for all particles. We first consider the weak and strong forces and then we discuss the long\/-\/range electromagnetism and gravity.

We recall that the algebraic operations which we usually do with words are
\begin{itemize}
\item writing them consecutively,
\item disrupting a word in between two letters %(actually, syllables) 
in accordance with the rules of hy\-phe\-n\-a\-ti\-on.
\end{itemize}
Notice that the first option prevails in frequency over the second whenever a sufficiently long fragment of text is already written; however, %the 
Nature first constructed the hy\-phe\-na\-ti\-on table for its glossary.

\section{The weak force}\noindent%
The defining property of a weak process by which it is recognised at once is an arbitrary combination of the following very unlikely events:
\begin{itemize}
\item a zero\/-\/length word, which means just one vertex in terms of paths, expands to a synonymic \emph{cycle} which is walked twice in opposite directions;
\item a path, being either a previously existing contour from the reaction's input or a newly\/-\/produced synonym of~$\langle1\rangle$, is torn.
\end{itemize}
Then the available collection of paths' fragments, which themselves may be not cyclic words but their separate letters or syllables, recombine and join the loose ends, forming new contours and thus creating new particles. Note that the output of a weak reaction is an anagram of %the 
letters belonging to the original text~--- which was extended with extra letters and their negations by using the add\/-\/subtract arithmetic trick.

\begin{example}
Let us consider the weak decay of a free neutron~$\mathsf{n}^0$; having spin $\pm\hbar/2$ and parity~$+1$ (i.e., the orientation\/-\/matching of the unknot's word and the orientation of the lattice is ``yes''), %the 
neutron's contour resembles the one drawn in Fig.~\ref{FigNeutron}(a).
\begin{figure}[htb]
{\unitlength=1mm
\centerline{\begin{picture}(75,70)(0,-8)
%%%%%%%%%%%%%%%%%%%%%%%%%%%%%%%%%%%%%%%%%% OLD %%%%%%%%%%%%%%%%%%
\put(0,34){\begin{picture}(0,0)
%%%%%%%%%%%%%%%%%%%%%%%%%%%%%%%%%%%%%%%%%%
\put(2,25){\begin{picture}(7,7)
\put(-2,0){\vector(1,0){7}}
\put(0,-2){\vector(0,1){7}}
\put(-1,-1){\vector(1,1){4}}
\put(5.5,-1){${\scriptstyle{1}}$}
\put(3.5,3){${\scriptstyle{2}}$}
\put(-1,4){\llap{${\scriptstyle{3}}$}}
\end{picture}
}
%%%%
\put(0,0){\begin{picture}(25,25)(-10,0)
\put(0,0){\circle*{1}}
\put(0,0){\vector(0,1){16}}
\put(0,16){\vector(1,1){8}}
\put(8,24){\vector(-1,0){4}}
\put(4,24){\line(0,-1){3}}
\put(4,19){\vector(0,-1){10}}
\qbezier(1,6)(4,9)(4,9)
%\put(1,5){\line(1,1){3}}%\put(4,8){\line(-1,-1){3}}
\put(-1,4){\vector(-1,-1){5}}
\put(-6,-1){\vector(1,0){18}}
\put(12,0){\circle*{1}}
\qbezier[12](12,0)(0,0)(0,0)
\put(2,0){\vector(-1,0){1}}
\put(4,1.5){939\,\textsf{MeV}}
\put(0,-8){$\mathsf{n}^0$}
\put(-5,10){\llap{(a)}}
\end{picture}
}
%%%%
\put(50,0){\begin{picture}(25,25)(-5,0)
\put(0,0){\circle{1}}
\put(-5,0){\line(1,0){4.5}}
\put(-0.5,0){\vector(0,1){20}}
\put(0,20){\circle*{1}}
\put(0.5,20){\vector(0,-1){19.5}}
\put(0.5,0){\vector(1,0){18.5}}
\put(20,-0.5){\circle*{1}}
\qbezier[25](19,-1.5)(-5,-1.5)(-5,-1.5)
\put(-4,-1.5){\vector(-1,0){1}}
%%%
\qbezier(0,20)(-12.5,20)(-12.5,27.5)
\qbezier(0,20)(-7.5,32.5)(-12.5,27.5)
\put(-2,20){\vector(1,0){1}}
\qbezier(0,20)(12.5,20)(12.5,27.5)
\qbezier(0,20)(7.5,32.5)(12.5,27.5)
\put(2,20){\vector(-1,0){1}}
\put(-13,22){\llap{$\BBS^{+1}$}}
\put(13,22){$\BBS^{-1}$}
%%%
\put(2.5,1.5){\circle*{1}}
\put(2,1.5){\vector(0,1){16.5}}
\put(2,18){\vector(1,-1){17}}
\put(19,1){\vector(-1,0){16}}
\put(2.5,2){\vector(1,0){14.5}}
\put(17,2){\vector(-1,1){14}}
\put(3,16){\vector(0,-1){14}}
\put(6,5){$\circlearrowright$}
\put(-6,10){\llap{(b)}}
\end{picture}
}
\end{picture}}%%%%%%%%%%%%%%%%%%%%%%%%%%%%%%%%%%%%%%%%%
%%%%
\put(10,0){\begin{picture}(25,25)
\put(0,0){\vector(1,0){8}}
\put(9,-0.5){\circle*{1}}
\qbezier[11](9,-1.5)(0,-1.5)(0,-1.5)
\put(1,-1.5){\vector(-1,0){1}}
\qbezier(9,0)(3,10)(9,10)
\qbezier(9,0)(15,10)(9,10)
\put(8,2){\vector(1,-2){1}}
\put(7,11){$\BBS^{+1}$}
\put(0,1){938}
\put(3,-8){$\mathsf{p}^{+}$}
%%%%
\put(13,0){\line(1,0){4}}
\put(13,-0.7){\line(0,1){1.4}}
\put(17,-0.7){\line(0,1){1.4}}
\put(13,1){0.8}
%%%%
\put(21,0){\begin{picture}(18,18)
\put(0,0){\circle*{1}}
\put(0,0){\line(3,5){9}}
\put(9,15){\circle*{1}}
\qbezier(9,15)(3,25)(9,25)
\qbezier(9,15)(15,25)(9,25)
\put(10,17){\vector(-1,-2){1}}
\put(14,22){$\BBS^{-1}$}
\put(8,13){\vector(1,2){0.5}}
\put(9,15){\line(3,-5){9}}
\put(17,2){\vector(1,-2){0.5}}
\put(18,0){\circle*{1}}
\qbezier[18](18,0)(0,0)(0,0)
\put(1,0){\vector(-1,0){0.5}}
\put(7.5,7){$\circlearrowright$}
\put(4,1){0.511}
\put(8,-8){$\mathsf{e}^{-}$}
\end{picture}
}
%%%%
\put(43,0){\begin{picture}(18,18)
\put(0,0){\circle*{1}}
\put(0,0){\line(3,5){9}}
\put(1,2){\vector(-1,-2){0.5}}
\put(9,15){\circle*{1}}
\put(9,15){\line(3,-5){9}}
\put(10,13){\vector(-1,2){0.5}}
\put(18,0){\circle*{1}}
\put(0,0){\vector(1,0){17.5}}
\put(7.5,5){$\circlearrowright$}
\put(8,-8){$\overline{\nu}_{\mathsf{e}}$}
\end{picture}
}
\put(-1,10){\llap{(c)}}
\end{picture}
}
%%%%%%%%%%%%%%%%%%%%%%%%%%%%%%%%%%%%%%%%%% OLD %%%%%%%%%%%%%%%%%%
\end{picture}
}
}\caption{Decay of free neutron.}\label{FigNeutron}
\end{figure}
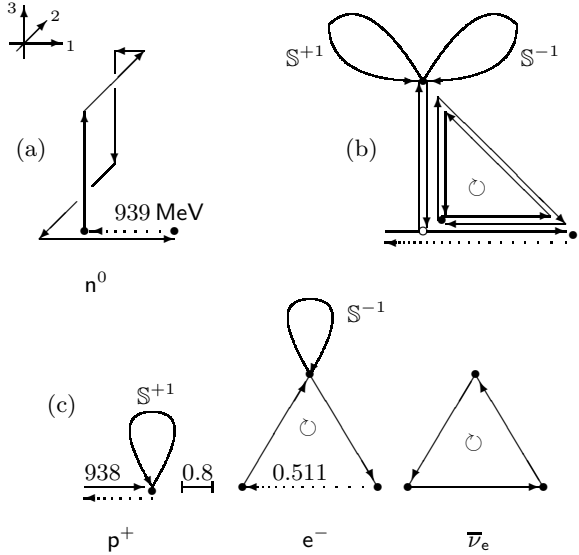
The primary channel (${}\approx99.9\%$) of the $\beta$-\/decay is
\begin{equation}\label{EqBetaDecay}
\mathsf{n}^0\longmapsto\mathsf{p}^{+}+\mathsf{e}^{-}+\overline{\nu}_{\mathsf{e}}+\langle0.8\,\text{\textsf{MeV}}\rangle.
\end{equation}
The reaction is energy\/-\/positive: the difference in mass between $\mathsf{n}^0$ and the proton~$\mathsf{p}^{+}$ exceeds the mass of electron; our previous argument in sec.~\ref{SecPart} (see Fig.~\ref{FigSorbonne}) shows that the electron (anti)neutrinos $\nu_{\mathsf{e}}$ and~$\overline{\nu}_{\mathsf{e}}$ are %most likely 
massless. The energy which is stored in the neutron's contracted edge(s) has the traditional choice of three fates: to remain stored in the proton's mass, to endow the electron with mass, or to be spent on the disruption of contours and exchange of edges; the rest is communicated to the three emitted particles as their kinetic energy.

The geometry of cross reactions refers to the same Fig.~\ref{FigNeutron}(b-c):
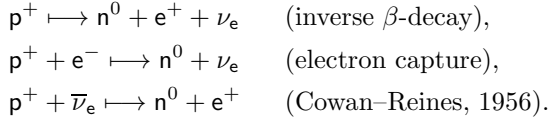
\begin{align*}
\mathsf{p}^{+}&\longmapsto\mathsf{n}^0+\mathsf{e}^{+}+\nu_{\mathsf{e}} & &\text{(inverse $\beta$-\/decay),}\\
\mathsf{p}^{+}&+\mathsf{e}^{-}\longmapsto\mathsf{n}^0+\nu_{\mathsf{e}} & &\text{(electron capture),}\\
\mathsf{p}^{+}&+\overline{\nu}_{\mathsf{e}}\longmapsto\mathsf{n}^0+\mathsf{e}^{+}
& &\text{(Cowan\/--\/Reines, 1956).}
\end{align*}
Another channel (${}\approx0.1\%$) of %a 
free neutron %'s 
decay is
\[
\mathsf{n}^0\longmapsto\mathsf{p}^{+}+\mathsf{e}^{-}+\overline{\nu}_{\mathsf{e}}+\gamma.
\]
This reaction inserts the null path $\langle1\rangle=\langle\mathbb{S}^1\mathbb{S}^{-1}\rangle$ in between the factors which assign the opposite charges to $\mathsf{p}^{+}$ and $\mathsf{e}^{-}$:
\[
\langle1\rangle=\langle\underbrace{\mathbb{S}^1}_{+e}\cdot\underbrace{\mathbb{S}^{-1}}_{-e}\rangle=\langle\underbrace{\mathbb{S}^1}_{\mathsf{p}^{+}}\cdot
 \underbrace{\mathbb{S}^1\mathbb{S}^{-1}}_{\gamma}\cdot
\underbrace{\mathbb{S}^{-1}}_{\mathsf{e}^{-}}\rangle.
\]
The proportion $0.1\%$ demonstrates how unwilling \textsc{sator Arepo} is to rotate twice not once around the circle in the compactified dimension (see Fig.~\ref{FigNeutron}(b)).
\end{example}

We conclude that a %the 
smooth complex $SU(2)$-\/gauge theory over the 
space\/-\/time viewed as a smooth manifold could %have no direct relation to 
be not a verbatim phrasing of %the 
Nature's prosaic effort in describing the weak interaction.

\section{The strong force}\label{SecInterStrong}\noindent%
The strong interaction does not beg, borrow, or steal the contours which did not belong to %the 
reaction's input but which could be created at the expense of %the 
zero\/-\/length words of energy; the output of a strong reaction looks much the same as the original text, only the order of sentences in it is mixed. For example, several protons and neutrons link to a bouquet or form a chain mail, that is, a nucleous. We emphasize that the strong interaction handles the already\/-\/existing contours (i.e., not just %the 
words encoded in the glossary via route instructions and edge contraction configurations).

The binary algebraic operation that creates the strong force is the multiplication~$\times$ for cyclic words; we described it in detail in~\cite{SQS11,Lorentz12} (see also~\cite{KontsevichCyclic}). Essentially, this is the standard unlock\/-\/and\/-\/join technique of the topological pair of pants~$S^1\times S^1\to S^1$. By definition, the value of operation~$\times$ at two paths is calculated in three steps:
\begin{enumerate}
\item both contours are unlocked at one vertex each;
\item the unlocked paths are transported along the lattice such that the two (un)locks coincide;
\item the loose ends of %the 
disrupted contours are recombined in such a way that the left\/-\/to\/-\/right order of reading all %the 
words is preserved.
\end{enumerate}
The structure~$\times$ takes the sum over all possible (or preferred) locations of the locks on each of the contours; if a certain summand is forbidden 
(recall that %e.g., if 
oriented spatial edges of the graph are viewed as fermions),
 %if the fermionic spatial edges do not allow the joined path to run the same edge in the same direction twice)
then it is omitted; the result is normalised by the actual number of contributing terms.

Notice %Note %
that the multiplication of %the 
particles' contours is com\-mu\-ta\-ti\-ve but not associative; indeed, the contours $|1\rangle$ and $|2\rangle$ are always adjacent in the product $\bigl(|1\rangle\times|2\rangle\bigr)\times|3\rangle$ but they can be separated by edges of $|3\rangle$ in some of the terms in $|1\rangle\times\bigl(|2\rangle\times|3\rangle\bigr)$.
%\marginpar{OK! ppn vs pnp}
Yet we recall that the strong interaction is not associative at the level of nuclear fusion and fission: e.g., the channel $\mathsf{p}^{+}\times\bigl(\mathsf{p}^{+}\times\mathsf{n}^{0}\bigr)\longmapsto{}_2^3\text{\textsf{He}}$ is not realised via $\bigl(\mathsf{p}^{+}\times\mathsf{p}^{+}\bigr)\times\mathsf{n}^{0}$ by a would\/-\/be intermediate two\/-\/proton~${}_2^2\text{\textsf{He}}$;
likewise, the equally possible (50\%\/-\/50\%) processes ${}_1^2\mathsf\mathsf{D}+{}_1^2\mathsf{D}\longmapsto{}_1^3\mathsf{T}+\mathsf{p}^{+}$ or $\longmapsto{}_2^3\text{\textsf{He}}+\mathsf{n}^0$ do not amount to a bare assembly of two protons and two neutrons.

%Nevertheless, to grasp the %match the experimentally observed 
%associativity by the triangle equation, %of the strong interaction,
%\[
%\bigl(|1\rangle\times|2\rangle\bigr)\times|3\rangle=
%|1\rangle\times\bigl(|2\rangle\times|3\rangle\bigr),
%\]
%   %in the course of the high\/-\/energy scattering <-> formation of atoms.
%one has to deform the commutative non\/-\/associative binary operation~$\times$ to the associative but not commutative star\/-\/product~$\star=\times+\text{const}\cdot\hbar\cdot\{\,,\,\}_{\text{Poisson}}+\overline{o}(\hbar)$, converting all the structures and operations at hand into power series in the Planck constant~$\hbar$. This could be done by a proper upgrade of the deformation quantisation 
%technique~\cite{KontsevichFormality,CattaneoFelderCMP2000}; 
%the %Hamiltonian formalism
%Poisson bracket~$\{\,,\,\}_{\text{Poisson}}$ of particles appears in the $\hbar^1$-\/slice of the full quantum picture.

%Note also that the expected property of the structure~$\star$ to be associative but not commutative implies the violation of the P-\/symmetry, which is (Left $\rightleftarrows$ Right) $\Longleftrightarrow$ ($|1\rangle\rightleftarrows|2\rangle$), by the strong force.

\smallskip
Finally, we recall that the full power of noncommutative %differential 
calculus %(which is needed as an intermediate tool for the approach of~\cite{KontsevichFormality}, see also~\cite{CattaneoFelderCMP2000}) 
is revealed by the introduction of %\textsl{jet spaces} $J^\infty(\pi^{\nC})$ over a 
noncommutative bundles~$\pi^{\nC}$ whose base is the noncommutative tangent bundle over space, i.e., the %quantum, 
\textsf{homeo}\/-\/class space %\/-\/time 
itself (see~\cite{Lorentz12}). Namely, let us consider \emph{open}, positive\/-\/length words (containing the mark\/-\/up of contracted edges) which we view as noncommutative fields over %the quantum 
space%\/-\/time
; the new structures are auxiliary in a description of the full quantum geometry of the strong force and may determine no particles existing as independent objects, yet they could be helpful. These fields are sections of the noncommutative bundle~$\pi^{\nC}$ over the \textsf{homeo}\/-\/class 
space. %\/-\/time. 
At each point of the base, a local basis of such fields extends the alphabet 
$\BBS%\vec{\mathrm{t}}
^{\pm1}$, $\vx_i^{\,\pm1}$ of generators of the lattice; not without insight we denote by $(\mathsf{u},\mathsf{d},\mathsf{s};\mathsf{c},\mathsf{t},\mathsf{b})$ the elements of such bases introduced pointwise.\footnote{As usual, the inverse elements $(\overline{\mathsf{u}},\overline{\mathsf{d}},\overline{\mathsf{s}};\overline{\mathsf{c}},\overline{\mathsf{t}},\overline{\mathsf{b}})$ mean the inversions $\BBS^{\pm1}\mapsto\BBS^{\mp1}$, $\vx_i^{\,\pm1}\mapsto\vx_i^{\,\mp1}$ and reading the summands backwards in each component of a section for~$\pi^{\nC}$.}
We foresee that the supports of such %those 
(anti)\/sections are finite in space in all inertial reference frames yet such supports do not amount to single points but contribute to %the 
construction of clouds of matter, that is, dimensionful particles.

\begin{cor}
Provided that the information encoding leptons is referred to one point in space at each instant of time, and under the hypothesis that the sections of~$\pi^{\nC}$ are piecewise\/-\/continuous with respect to the continuous limits of space tilings, the leptons may not consist of 
  %finitely many 
such auxiliary building blocks.
\end{cor}

\begin{rem}
Because the auxiliary blocks $(\mathsf{u},\mathsf{d},\mathsf{s};\mathsf{c},\mathsf{t},\mathsf{b})$ are introduced to encode formal sums of paths which are {open}, they in practice can not be isolated physically and registered as objectively existing particles. (The same argument applies now, long after the Big Bang, to each generator~$\vx_i$ of a %n aperiodic affine 
lattice in space.) %~$\mathbb{E}^3$.)
\end{rem}

\section{Electromagnetism}\noindent%
Now we address the effective long\/-\/range interactions: electromagnetism first and then gravity. It must be noted that both concepts involve the same idea of %the 
edge contraction (in disguise, formation of tadpoles), that is, a prescription that the endpoints of an edge become one vertex. (The difference between the two concepts is that a spatial edge vanishes altogether, not forming a loop, whereas the generators $\mathbb{S}^{\pm1}$ %or~$\vec{\mathrm{t}}^{\,\pm1}$
are tadpoles for granted.) Therefore, it is logical to expect that %the
macroscopic properties of the two forces are much alike in their classical description: a charged massive point locally produces the Newtonian electric and gravitational potentials inverse\/-\/proportional to distance in space.

\begin{rem}
The count of electric charge, which is by definition the difference
$\sharp\mathbb{S}^{+1}-\sharp\mathbb{S}^{-1}$
of the numbers of loops that a path winds in the fourth, tadpole dimension, 
does not interfere with disruptions or rearrangements of %the 
path's spatial edges. Consequently, the electric charge is conserved but in %all 
other respects it does not influence %the 
weak or strong processes.
\end{rem}

The large\/-\/scale alikeness of the long\/-\/range forces is readily seen from the coding of electron on the equilateral triangular lattice, see Fig.~\ref{FigElectron}:
%%%%%%%%%%%%%%%
%%% Fig. 6
%%%%%%%%%%%%%%%
the particle consists of indivisible mass and indivisible charge which are held close to each other but always stay separated by emptiness; the information about %the 
particle's closed contour does not take shape of a faintly shimmering rope or cord.
\begin{figure}[htb]
{\unitlength=1mm
\centerline{
\begin{picture}(50,16)(-12.5,-8)
\put(0,0){\circle{1}}
\qbezier(1,0.5)(12.5,10)(24,0.5)
\put(23,-1){\vector(2,1){1}}
\put(25,0){\circle*{1}}
\qbezier(1,-0.5)(12.5,-10)(24,-0.5)
\put(2,1){\vector(-2,-1){1}}
\put(11,-1){$\circlearrowleft$}
\put(9,-8){$\longleftrightarrow$}
%%%%
\qbezier[16](-1,-0.5)(-12.5,-7.5)(-12.5,0)
\qbezier[16](-1,0.5)(-12.5,7.5)(-12.5,0)
\put(-2,-1){\vector(2,1){1}}
\put(-13.5,0){\llap{$m$}} %\approx0.511\text{\,\textsf{MeV}}/c^2$}}
%%%%
\put(25,0){\begin{picture}(12.5,7.5)
\qbezier(1,0.5)(12.5,7.5)(12.5,0)
\qbezier(1,-0.5)(12.5,-7.5)(12.5,0)
\put(2,-1){\vector(-2,1){1}}
\put(14,0){$\BBS^{-1}$}
\end{picture}
}
\end{picture}
}
}\caption{The electron~$\mathsf{e}^{-}$.}\label{FigElectron}
\end{figure}
Note that the contraction of the edge which endows electron with mass creates the asymmetry of space surrounding the point where the mass is located; on the other hand, the formation of negative charge ${-}e$ by the tadpole~$\mathbb{S}^{-1}$ is fully symmetric with respect to space.
  %Magneton; quantum oscillator.
%\marginpar{Edit}

\begin{rem}\label{RemCompton}
It is logical also that, whenever accelerating in electromagnetic field, the electron {radiates}. Indeed, Lorentz' force acting on its charge, the electron becomes an oscillator in which space itself plays the r\^o\-le of elastic spring between the point charge and the point mass; the oscillations then amount to periodic elastic deformations of the space\/-\/time structure, pulling or slowing the mass as it retards or overtakes the charge.
\end{rem}

In conclusion, electromagnetism is a very famous example of cyclic\/-\/invariant theory; outside point particles, its gauge description could be \emph{exact}.

\section{Gravity: the Big Bang logistics}%
\noindent(We owe this term to Yu.~I.~Manin who coined~it.)
%\footnote{; let us also recall that the ordering of the Universe in small boxes and events which led to the Big Bang are depicted in the Russian edition of~\cite{Metaphor} on p.~268.}

\smallskip
Let us attempt to track logically the scenario of events in the early Universe, taking into account the assertions which we have made so far.

1. Initially, the Universe consisted (according to its topology) of only one point; in other words, %which is encoded by the statement that 
the CW-\/complex was fully contracted, i.e., %thus 
formed by only one vertex and no edges. It is possible that the initial point was also assigned an extra number that indicated the energy surplus over its store in the contracted edges.

2. The fully contracted lattice was released from hold and its edges began to decontract; each event of edge decontraction released energy which took shape of the zero\/-\/length word~$\langle1\rangle$. The time started; it first amounted to the count of decontraction events and derivative events of reconfigurations in the lattice defects' portrait; still no particles were formed yet and there was no light.

A decontraction of the first spatial edge with tadpoles at each of its ends created the possibility of existence of light viewed as the automaton that propagates the null photon's path $\langle\mathbb{S}_m^{\pm1}\cdot\mathbb{S}_n^{\mp1}\rangle$ from a vertex to its neighbour; this does not imply that the ready\/-\/to\/-\/work automaton did actually start to work at once.
The decontraction of the first triangle did create the first electron neutrino (see Fig.~\ref{FigSorbonne}); we think that it is scholastic to debate whether the first photon preceded the first neutrino or \textit{vice versa}.
For the first tetrahedron to decontract, a coin was thrown and this determined whether that was a Left or Right tetrahedron (see Fig.~\ref{FigTetra}); space was thus oriented and Nature chose Right.

The dimension of the CW-\/complex became positive, and the initial point split to a set of points (depending on the convention about \textsf{homeo}\/-\/class continuous space or quantum space as a lattice, to continua or to finite sets through a finite number of decontraction events, respectively); space began to expand.

3. While the edges kept on decontracting in the replicas of adjacency table for each newly\/-\/produced copy of the original vertex, such events became independent and uncorrelated. Because of this, the values of noncommutative Ricci and scalar curvature were (almost) random at points of the early Universe.\footnote{The worst idea here would be that of {averaging}, see~\cite{DDSokolov}.} However, let us recall from~\cite{DDSokolov} that in this case the Jacobi field connecting two infinitesimally close geodesics issued from a point grew exponentially~--- yet, paradoxically, for almost certain there appeared an arbitrarily large number of those geodesics' focus points, which resulted in clashes and information exchange. (Here we use the assumption of space's continuity and use light to introduce the smooth structure of space at the expense of infinite energy.) The Universe experienced the \emph{inflation}.

4. It took certain time for space to expand and reach a configuration with %a 
relatively small fraction of defect edges that %which 
remained contracted in large finite neighbourhoods of many points. (Because the Universe continues expanding {now}, we may not refer to all %of 
its points but operate with sufficiently spacious regions.) Simultaneously, those lattice defects could assemble due to the reasons of entropy. However, the preceding decontraction of a major part of edges in such %those 
domains released a colossal amount of vacuum energy; it shaped into a scalar field over space.

5. Although the initial point of the Universe split (its descendants still continue splitting at the outer periphery of the Universe, sending us photons and neutrino flows; throughout Cosmos, the edges between its descendants
continue splitting and so contribute to the validity of Hubble's law), there re\-ma\-i\-ned or there re\/-\/appeared sets of ``conservative\/-\/min\-ded'' vertices in the lattice which proclaimed them\-sel\-ves one\/-\/point\/-\/forever and merged the adja\-cen\-cy tables~--- but their neighbours refused to join the coalition. This created the singularities of the first generation of black holes. (Those were the ti\-mes when a lack of curvature necessary to form the event horizon was out of dispute.) The eldest black holes produced considerable irregularities of the space\/(-time) geometry, which triggered the formation of matter from an otherwise still meta\/-\/stable state of the Universe already filled with neutrinos and photons (cf.~\cite{Hawking}).

6. The excess of vacuum energy, the decontractions which created massless neutrino cycles as in Fig.~\ref{FigSorbonne}, photons, and
  %synonymic to the trivial word~$\langle1\rangle$, 
possible entropy\/-\/based gradient flows of the dark matter caused the formation of elementary particles via weak processes of contour disruption and reconfiguration (in particular, relatively near --in cosmic sense-- to black holes). The Left\/-\/Right asymmetry, which had been built into the Universe since the decontraction of the first tetrahedron, implied the domination of matter over antimatter in the course of particle creation. The add\/-\/subtract mechanism for creating, distributing, and counting the electric charge kept the initially pathless and hence chargeless Universe neutral \textit{en masse}.

7. The world became what we know it now; only a small part of its mass and energy shaped into particles. Nevertheless, those were enough to form galaxies around the eldest black holes which called that matter from unbeing. The start of chain reactions of nuclear fusion and ignition of the first star heralded the end of the beginning in the history of the Universe.

\bigskip
%\section{Conclusion}%
\noindent\textbf{Conclusion}\\[4pt]%
In this review, preceded by~\cite{Protaras12}, we %have 
outlined a possible axiomatic quantum picture of %the 
fundamental interactions in quantum space and time; we hope that it will help us to resolve a part of known difficulties or at least offer us a good reformulation of paradoxes in the existing paradigm.

\smallskip
We have shown that the chosen set of postulates implies the following statements:

\noindent%
1. At (sub-)\/Planck scale, gauge theory is insufficient for a description of the interactions; one should use a theory that does not appeal to the locally\/-\/linear, \textsf{diffeo}\/-\/structure of the space\/-\/time but operate with  geometry of the Universe at the topological, \textsf{homeo}\/-\/class level.

\noindent%
2. Vacuum, i.e., a domain in space which is known to contain no particles of any kind, can have mass\/-\/energy and, via its curvature, nonetheless produce gravity force; moreover, such vacuum does contain a store of energy which can be released and transmute into (anti)\/matter.

\noindent%
3. The (anti)\/matter is the meaning of information which is stored in space (specifically, within its topology and the loops paved through its 
\textsf{homeo}\/-\/realisation).

\noindent%
4. Having begun to expand from its initial state with the topology $\mathcal{T}=\{\varnothing,\text{Universe}\}$, the early Universe experienced an inflation phase of exponential growth.

\noindent%
5. The Universe is electrically neutral. In the quantum world, electromagnetism does contribute to the processing of information with the quantum number \emph{charge} but plays no dominant r\^o\-le in the interactions and decays (e.g., protons and neutrons form the nuclei of atoms). Outside the particles and under the \textit{ad hoc} assumption of differentiability for the gauge transformations, the $U(1)$-\/gauge model is {exact}.

\smallskip
We %also 
%logically 
argued in favour of the following possibilities:

\noindent%
6. The CP-\/symmetry violation in weak processes is a consequence of the Left~$\neq$ Right asymmetry which has been being built into the Universe since it became oriented; matter prevails now over antimatter. The masses (whenever both are nonzero) of the respective (anti)\/particles can be slightly unequal.

\noindent%
7. %The mass of electron's (anti)\/neutrino is identically zero.
The cosmic microwave background radiation is an immanent property of space itself, so it can not be shielded altogether by using any macroscopic medium.

\noindent%
8. The polarised antiphotons are cross\/-\/identical to the respective photons.

\noindent%
9. The weak processes are much less likely to occur than the strong ones.

\noindent%
10. A decay of the unstable leptons $\mu^{\pm}$ and $\tau^{\pm}$ and the (anti)neutrino oscillations $\nu_\tau\leftrightarrow\nu_\mu\leftrightarrow\nu_{\mathsf{e}}$ are second order phase transitions.

\smallskip
In conclusion, it is quite remarkable that matter~\emph{is}, even if it constitutes only about 16\% of the should\/-\/be mass (and possibly 4\% of the mass\/-\/energy) of the %visible 
Universe. The main store of mass is contained in the dark matter, i.e., in %the 
graphs' contracted edges where no nontrivial contours run, and there is a huge %tremendous 
store of the vacuum energy in zero\/-\/length words or %and 
their synonyms.

We conjecture that the organisation of (anti)\/matter in galaxies around very massive objects, as we observe it now, summarises the history of the Universe itself. Namely, we view the singularities of black holes in the centres of galaxies as the eldest remnants of the Universe's One Point %in~$\mathcal{T}$
that did not expand to continua but served, and do so presently, as those inhomogeneities for the near\/-\/by regions of space which catalysed the creation of matter, i.e., release of contracted edges and formation of cycles (the principle %being 
similar to boiling or condensation on admixtures). Thus, the presence of %a 
black hole in each galaxy's %the of  
centre is not their fatality but %the 
primordial blessing for their existence that called from unbeing the matter in them and then shaped it. For the same reason, it is unlikely that there are many lone stars outside galaxies.

Likewise, we expect that %the 
giant voids in the large\/-\/scale structure of the Universe, the voids bounded by %the 
walls formed by galaxies, are relatively poor in black holes and therefore are presently filled with the still dark matter and vacuum energy (however, in meta\/-\/stable state). This is why the voids (or each galaxy's halo)
look so empty even if they are so massive. Nevertheless, the superclusters, filaments, and walls in %the 
large\/-\/scale organisation of the Cosmos stem from a %the
topological configuration of vertices, edges, and cells in Planck\/-\/scale 
   %tiniest 
neighbourhoods of all its points.

\vskip3mm %\ack
The author thanks the Organizing committee of International conference 
`Symmetries of discrete systems \& processes' 
(July 15--19, 2013 at CVUT D\v{e}\v{c}\'{\i}n, Czech Republic)
and the organizers of
V~Conference of young scientists `Problems of Theoretical Physics' 
(held December 24--27, 2013 at Bogolyubov~ITP NAS in Kiev, Ukraine)
  %International conference `Quantum groups \& quantum integrable systems' (June 18\/--\/21, 2013; Bogolyubov ITP, Kiev, Ukraine) 
for stimulating discussions and partial financial support.
This research was supported in part 
by %NWO~VENI grant~639.031.623 (Utrecht) and 
JBI~RUG project~103511 (Groningen). 
A~part of this research was done while the author was visiting at 
the $\smash{\text{IH\'ES}}$ (Bures\/-\/sur\/-\/Yvette); 
the financial support and hospitality of this institution are gratefully acknowledged.

%\vspace*{-5mm} 
%\rezume{%
%\mbox{ }
%}

\end{document}